\documentclass[pdflatex,sn-mathphys,Numbered,iicol]{sn-jnl}

\usepackage{nicefrac}
\usepackage[draft,todonotes={textwidth=1.5cm,tickmarkheight=0cm}]{changes} 

\usepackage{multirow}
\usepackage{siunitx} %degree symbol, etc.
\usepackage{gensymb}
\usepackage{amssymb,amsbsy,amsmath,amsfonts}
\usepackage{color,graphicx}
\usepackage{multicol}
\usepackage{xcolor}
\usepackage{epsf,epsfig,float,latexsym,amsthm,fancyhdr,rotating}
\usepackage{graphics,psfrag}
\usepackage{slashed}
\usepackage{comment}
\usepackage{lipsum}

\usepackage[labelfont=bf,font+=smaller,justification=centerlast]{caption}

\usepackage{upgreek}
\hypersetup{colorlinks,citecolor=blue,linktoc=all,linkcolor=cyan,urlcolor=blue}

\usepackage{epsfig}
\usepackage{subfig}
\usepackage{sidecap}
\usepackage{keyval}
\usepackage{wrapfig}

\usepackage{dcolumn}
\usepackage{bm}

\usepackage{textcomp}

\usepackage{setspace}
 
\usepackage[normalem]{ulem}

\usepackage{cleveref}
\crefname{figure}{Fig.\@}{Figs.\@}
\crefname{equation}{Eq.\@}{Eqs.\@}
\crefname{section}{Sec.\@}{Secs.\@}
\crefname{table}{Table}{Tables}

\def\beq{\begin{equation}}
\def\eeq{\end{equation}}
\def\bea{\begin{eqnarray}}
\def\eea{\end{eqnarray}}
\def\beqa{\begin{equation}\begin{array}{l}}
\def\eeqa{\end{array}\end{equation}}

\def\barr{\left(\begin{array}{c}}
\def\earr{\end{array}\right)}
\def\bmat{\left(\begin{array}{cc}}
\def\emat{\end{array}\right)}

\def\3d{3-D}

\definecolor{darkred}{rgb}{0.9, 0.0, 0.0}

\title{The Two-Photon Exchange Experiment at DESY}

\author[1]{\fnm{R.}~\sur{Alarcon}}
\author[2]{\fnm{R.}~\sur{Beck}}
\author[3,4]{\fnm{J.~C.}~\sur{Bernauer}}
\author[5]{\fnm{M.}~\sur{Broering}}
\author[6]{\fnm{A.}~\sur{Christopher}} %added by Michael
\author*[3,5]{\fnm{E.~W.}~\sur{Cline}}\email{ewcline@mit.edu}
\author[6]{\fnm{S.}~\sur{Dhital}}
\author[6]{\fnm{B.}~\sur{Dongwi}}
\author[6]{\fnm{I.}~\sur{Fernando}}
\author[7]{\fnm{M.}~\sur{Finger}}
\author[7]{\fnm{M.}~\sur{Finger~Jr.}}
\author[8]{\fnm{I.}~\sur{Fri{\v s}{\v c}i\'c}}
\author[6]{\fnm{T.}~\sur{Gautam}}
\author[9]{\fnm{G.~N.}~\sur{Grauvogel}}
\author[5]{\fnm{D.~K.}~\sur{Hasell}}
\author[5]{\fnm{O.}~\sur{Hen}}
\author[10]{\fnm{T.}~\sur{Horn}}
\author[5]{\fnm{E.}~\sur{Ihloff}}
\author[5]{\fnm{R.}~\sur{Johnston}}
\author[5]{\fnm{J.}~\sur{Kelsey}}
\author[6]{\fnm{M.}~\sur{Kohl}}
\author[5,9]{\fnm{T.}~\sur{Kutz}}
\author*[11]{\fnm{I.}~\sur{Lavrukhin}}\email{ievgen@umich.edu}
\author[5]{\fnm{S.}~\sur{Lee}}
\author[11]{\fnm{W.}~\sur{Lorenzon}}
\author[11]{\fnm{S.}~\sur{Lunkenheimer}}
\author[12]{\fnm{F.}~\sur{Maas}}
\author[5]{\fnm{R.~G.}~\sur{Milner}}
\author[5]{\fnm{P.}~\sur{Moran}}
\author[6]{\fnm{J.}~\sur{Nazeer}}
\author[6]{\fnm{T.}~\sur{Patel}}
\author[6]{\fnm{M.}~\sur{Rathnayake}}
\author[11]{\fnm{R.}~\sur{Raymond}}
\author[5]{\fnm{R.~P.}~\sur{Redwine}}
\author[9]{\fnm{A.}~\sur{Schmidt}}
\author[13]{\fnm{U.}~\sur{Schneekloth}}
\author[14]{\fnm{D.}~\sur{Sokhan}}
\author[6]{\fnm{M.}~\sur{Suresh}}
\author[5]{\fnm{C.}~\sur{Vidal}}
\author[11]{\fnm{Z.}~\sur{Yang}}

\affil[1]{\orgdiv{Department of Physics and Astronomy}, \orgname{Arizona State University}, \orgaddress{\city{Tempe},~\state{AZ},~\country{USA}}}
\affil[2]{\orgdiv{Department of Physics and Astronomy}, \orgname{Friedrich Wilhelms Universit\"at}, \orgaddress{\city{Bonn},~\country{Germany}}}
\affil[3]{\orgdiv{Center for Frontiers in Nuclear Science}, \orgname{Stony Brook University}, \orgaddress{\city{Stony Brook},~\state{NY},~\country{USA}}}
\affil[4]{\orgdiv{Riken BNL Research Center}, \orgname{Brookhaven National Laboratory}, \orgaddress{\city{Upton},~\state{NY},~\country{USA}}}
\affil[5]{\orgdiv{Laboratory for Nuclear Science}, \orgname{Massachusetts Institute of Technology}, \orgaddress{\city{Cambridge},~\state{MA},~\country{USA}}}
\affil[6]{\orgdiv{Department of Physics and Astronomy}, \orgname{Hampton University}, \orgaddress{\city{Hampton},~\state{VA},~\country{USA}}}
\affil[7]{\orgdiv{Department of Low Temperature Physics}, \orgname{Charles University}, \orgaddress{\city{Prague},~\country{Czech Republic}}}
\affil[8]{\orgdiv{Department of Physics, Faculty of Science}, \orgname{University of Zagreb}, \orgaddress{\city{Zagreb},~\country{Croatia}}}
\affil[9]{\orgdiv{Department of Physics and Astronomy}, \orgname{The George Washington University}, \orgaddress{\city{Washington, DC},~\country{USA}}}
\affil[10]{\orgdiv{Department of Physics}, \orgname{Catholic University of America}, \orgaddress{\city{Washington, DC},~\country{USA}}}
\affil[11]{\orgdiv{Randall Laboratory of Physics}, \orgname{University of Michigan},~\orgaddress{ Ann Arbor},~\state{MI},~\country{USA}}
\affil[12]{\orgdiv{Institute of Nuclear Physics}, \orgname{Johannes Gutenberg Universit\"at}, \orgaddress{\city{Mainz},~\country{Germany}}}
\affil[13]{\orgname{Deutsches Elektronen-Synchrotron DESY}, \orgaddress{\country{Germany}}}
\affil[14]{\orgdiv{Department of Physics and Astronomy}, \orgname{University of Glasgow}, \orgaddress{\city{Glasgow},~\country{Scotland}}}

\abstract{
    We propose a new measurement of the ratio of positron-proton to electron-proton elastic scattering at DESY. The purpose is to determine the contributions beyond single-photon exchange, which are essential for the Quantum Electrodynamic (QED) description of the most fundamental process in hadronic physics. By utilizing a 20 cm long liquid hydrogen target in conjunction with the extracted beam from the DESY synchrotron, we can achieve an average luminosity of $2.12\times10^{35}$~cm$^{-2}\cdot$s$^{-1}\cdot$sr$^{-1}$~($\approx200$ times the luminosity achieved by OLYMPUS). The proposed TPEX experiment entails a commissioning run at 2~GeV, followed by measurements at 3~GeV, thereby providing new data up to $Q^2=4.6$~(GeV/$c$)$^2$ (twice the range of current measurements). We present and discuss the proposed experimental setup, run plan, and expectations.
    }

\begin{document}

\maketitle

%***+****1****+****2****+****3****+****4****+****5****+****6****+****7****+****8

\section{Introduction}
\label{sec:intro}

Elastic lepton-proton scattering is a fundamental process that should be well described by QED.  Understanding this interaction is important to the scientific programs at FAIR, Jefferson Lab, and the future electron-ion collider (EIC) planned for Brookhaven.  It is described theoretically in the Standard Model by a perturbative expansion in $\alpha=\frac{1}{137}$ with radiative corrections. 

For over half a century it has been assumed that the leading single-photon exchange term adequately describes the scattering process and that higher-order terms are negligible.  However, recent experiments at Jefferson Lab have been widely interpreted as evidence that higher order terms are significant in elastic electron-proton scattering and must be included to correctly extract the proton elastic form factors.  Recent experiments, including the OLYMPUS experiment at DESY, show little evidence for significant contributions beyond single photon exchange up to $Q^2\approx2.3$~(GeV/c)$^2$.

To better understand the QED expansion, it is crucial to experimentally study it at higher $Q^2$ by comparing the positron and electron scattering cross-sections. This comparison can help determine the contribution of higher-order terms that are not typically included in radiative corrections.

\begin{figure}[!ht]
  \centering
  \includegraphics[width=0.46\textwidth, clip] {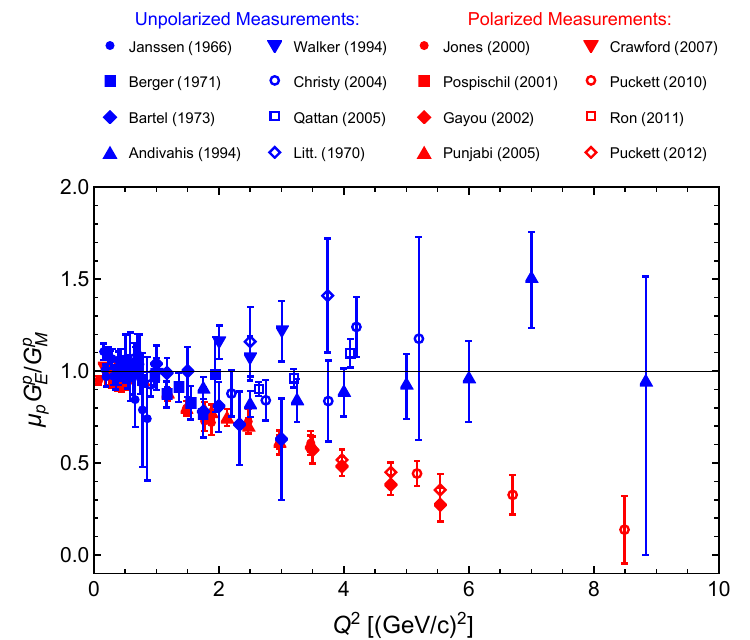}
  \caption{ Proton form factor ratio measured using unpolarized~\cite{Janssens:1965kd, Berger:1971kr, Litt:1969my, Bartel:1973rf, Andivahis:1994rq, Walker:1993vj, Christy:2004rc, Qattan:2004ht} (blue) and polarized~\cite{Jones:1999rz, Pospischil:2001pp, Gayou:2001qt, Punjabi:2005wq, Crawford:2007dl, Puckett:2010ac, Ron:2011rd, Puckett:2011xg} (red) techniques.}
  \label{fig:ratio}
\end{figure}

The proton form factors, $G_{E}^{p}$ and $G_{M}^{p}$, have
historically been envisaged as very similar and are often modeled by the same dipole form factor.  Measurements over the past 50~years using the unpolarized Rosenbluth separation technique yielded a ratio, $\mu^{p}\,G_{E}^{p}/G_{M}^{p}$, close to unity over a broad range in $Q^{2}$ shown by the blue data points in Figure~\ref{fig:ratio}.  This supported the idea that $G_{E}^{p}$ and $G_{M}^{p}$ are similar. However, recent measurements using polarization techniques revealed a completely different picture with the ratio decreasing rapidly with increasing $Q^{2}$ as shown by the red data points in Figure~\ref{fig:ratio}.

The most commonly proposed explanation for this discrepancy is ``hard'' two-photon exchange contributions beyond the standard radiative corrections to one-photon exchange.

\begin{figure}[!ht]
  \centering
  \includegraphics[width=0.4\textwidth, viewport=32 356 563 508, clip ]
    {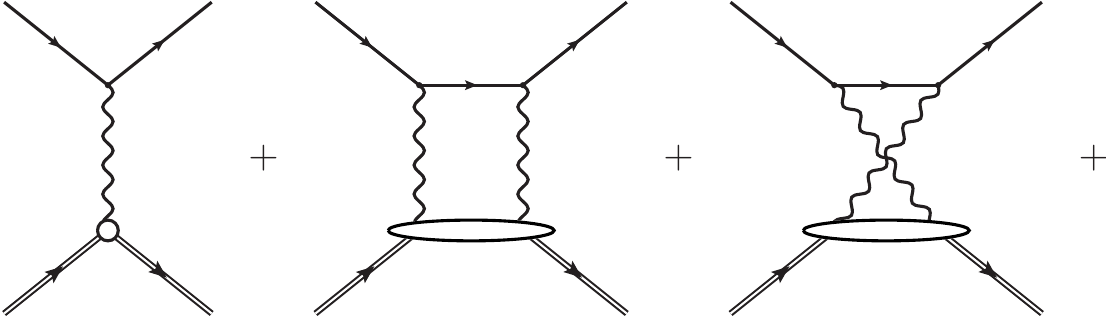}
  \caption{Feynman diagrams for one- and two-photon exchange. Further
    diagrams for bremsstrahlung, vertex, self-energy, and vacuum
    polarization radiative corrections are not shown but must also be
    included in calculations.}
  \label{fig:epLine1}
\end{figure}

Two-photon exchange, TPE, (see Figure~\ref{fig:epLine1}) is generally included as part of the radiative corrections when analyzing electron-proton scattering. However, it is usually only included in the ``soft'' limit where one of the two photons, in the diagrams with two photons, is
assumed to carry negligible momentum and the intermediate hadronic state remains a proton. To include ``hard'' two-photon exchange, a model for the off-shell, intermediate hadronic state must also be included, making the calculations difficult and model dependent.

In the Born or single photon exchange approximation the elastic scattering cross section for leptons from protons is given by the reduced Rosenbluth cross section,

\begin{equation}
    \frac{d\sigma_{e^\pm p}}{d\Omega}={\frac{d\sigma}{d\Omega}}_{Mott}
    \frac{\tau {G_M^p}^2 +\epsilon {G_E^p}^2}{\epsilon(1+\tau)},
\end{equation}
where: $\tau=\frac{Q^2}{4 M_p^2}$ and $\epsilon=(1 +2(1+\tau)
\tan^2{\frac{\theta_l}{2}} )^{-1}$.

To measure the ``hard'' two-photon contribution, one can measure the ratio $R_{2\gamma}=\sigma_{e^{+}p}/\sigma_{e^{-}p}$ at different values of $Q^{2}$ and $\epsilon$. Note, the interference terms between one- and two-photon exchange change sign between positron and electron scattering and this cross section ratio provides a measure of the
two-photon exchange contribution.

\begin{figure}[!ht]
  \centering
  \includegraphics[width=0.46\textwidth, clip]{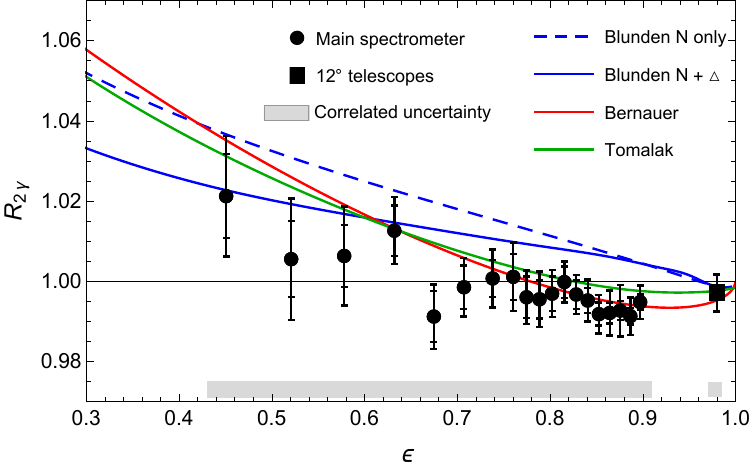}
  \caption{OLYMPUS results for $R_{2\gamma}$ as a function of $\epsilon$.  Inner error bars are statistical while the outer error bars include uncorrelated systematic uncertainties added in quadrature.  The gray band represents the correlated systematic uncertainty.}
  \label{fig:OLYMPUS}
\end{figure}

The results from the OLYMPUS experiment~\cite{Henderson:2016dea} are shown in Figure~\ref{fig:OLYMPUS} together with various calculations.

The deviation of the results from unity are small, on the order of 1\%, and are below unity at large $\epsilon$ and rising with decreasing $\epsilon$.  The dispersive calculations of Blunden~\cite{Blunden:2017nby} are systematically above the OLYMPUS results in this energy regime.  The results below unity cannot be explained by current QED calculations. The phenomenological prediction
from Bernauer~\cite{Bernauer:2013tpr} and the subtractive dispersion calculation from Tomalak~\cite{Tomalak:2014sva} are in better agreement with the OLYMPUS results but appear to rise too quickly as $\epsilon$ decreases.  There is some indication that TPE increases with decreasing $\epsilon$ or increasing $Q^{2}$, suggesting that a significant ``hard'' two-photon contribution might be present at lower $\epsilon$ or higher $Q^{2}$.

Two other experiments, VEPP-3~\cite{Rachek:2014fam} and
CLAS~\cite{Adikaram:2014ykv}, also measured the ``hard'' two-photon exchange contribution to electron-proton elastic scattering.

\begin{figure}[!ht]
  \centering
  \begin{subfloat}[]{\label{fig:diffblunden}
    \includegraphics[width=0.45\textwidth, clip]
  {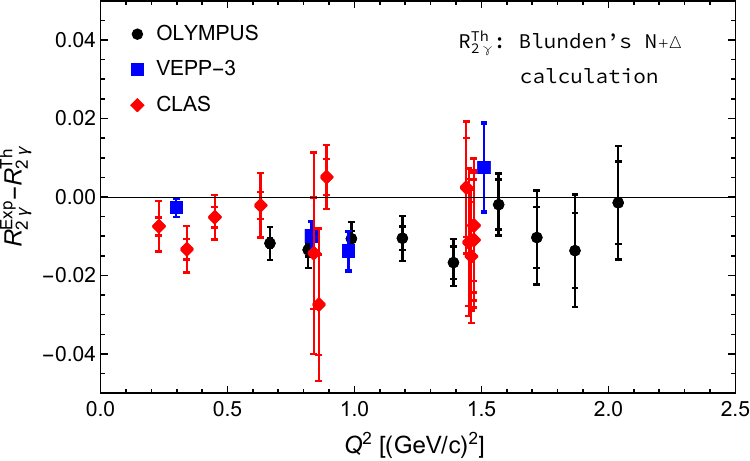}}
  \end{subfloat}
  \hfill
  \begin{subfloat}[]{\label{fig:diffbernauer}
    \includegraphics[width=0.45\textwidth, clip]
  {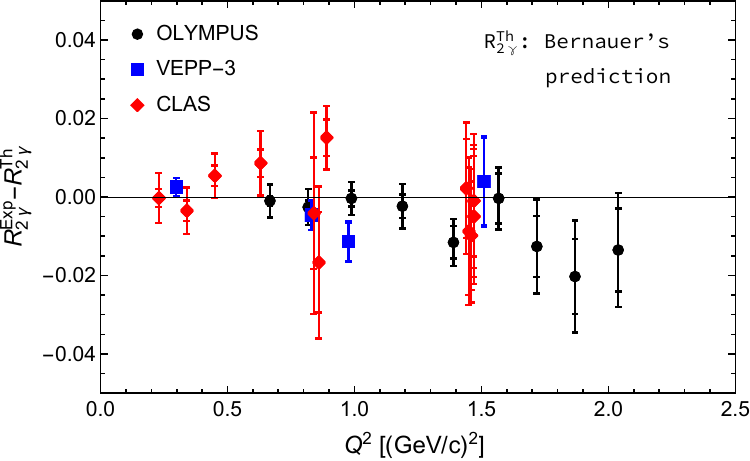}}
  \end{subfloat}
  \caption{Difference between the results from the three recent experiments and (a) Blunden's N+$\Delta$ calculation and (b) Bernauer's prediction.}.
\end{figure}

It is difficult to compare the results from the three experiments directly since the measurements are at different points in the $(\epsilon, Q^2)$ plane.  To partially account for this, we can compare all the two-photon exchange results by taking the difference
with respect to a selected calculation evaluated at the correct $(\epsilon, Q^2)$ for each data point.  This is shown
in Figure~\ref{fig:diffblunden} for Blunden's calculation and in Figure~\ref{fig:diffbernauer} for Bernauer's phenomenological prediction, plotted versus $Q^2$.  In these views, the results from the three experiments are shown to be in reasonable agreement supporting the previous conclusions.
 
The results from the three TPE experiments are all below
$Q^2=2.3$~(GeV/c)$^2$.  In this regime the discrepancy in the form factor ratios is not obvious, so the small ``hard'' TPE contribution measured is consistent with the measured form factor ratios.  The suggested slope with $\epsilon$ indicates TPE may be important at smaller $\epsilon$ or higher $Q^2$.  But, since this slope appears to deviate from Bernauer's phenomenological prediction, which fits the observed discrepancy, it may also suggest that ``hard'' TPE, while contributing, may not explain all of the observed form factor discrepancy.

\begin{figure}[!ht]
  \centering
  \includegraphics[width=0.46\textwidth] {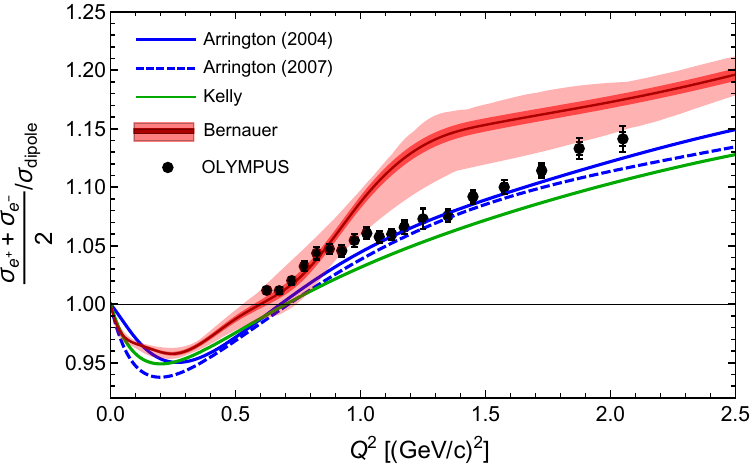}
  \caption{The charge-averaged yield for elastic lepton-proton scattering from the OLYMPUS experiment~\cite{Bernauer:2020vue}.}
  \label{fig:yield}
\end{figure}

Recently, the OLYMPUS data has also been analyzed to determine the charge-averaged yield for elastic lepton-proton scattering~\cite{Bernauer:2020vue}.  The result is shown in Figure~\ref{fig:yield}.

This measurement is insensitive to any charge-odd radiative
corrections including ``hard'' two-photon exchange and thus provides a better measure of the proton form factors.  The data shown covers an important range of $Q^2$ where the $G_M$ form factor changes slope. The calculations by Kelly~\cite{Kelly:2004hm} and Arrington~\cite{Arrington:2003ck,Arrington:2007ux} appear to be in better agreement with the data, but Bernauer's global fit~\cite{Bernauer:2013tpr} should be redone to incorporate all the OLYMPUS data.

\begin{figure}[htbp]
  \centering
  \includegraphics[width=0.46\textwidth] {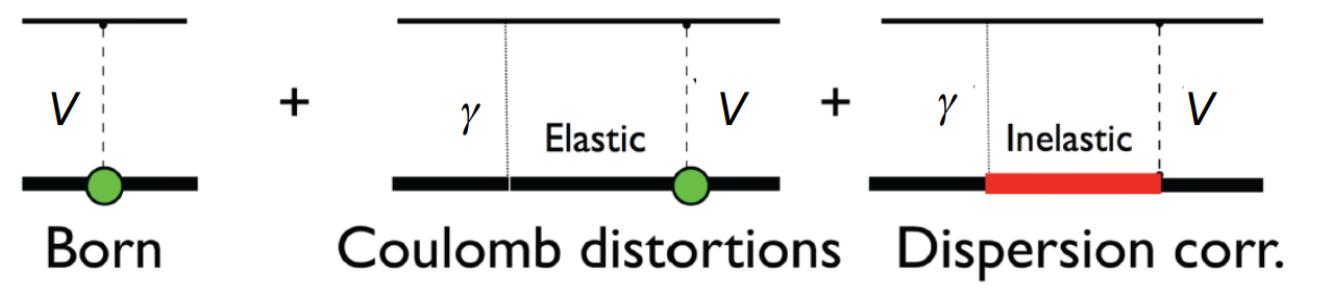}
  \caption{More general electroweak box diagram (where $V = Z^0, W^\pm, {\rm or}\ \gamma$) that is important in many fundamental nuclear physics processes.}
  \label{fig:ewbox}
\end{figure}

The two-photon exchange diagram in the QED expansion for electron scattering is an example of the more generic electroweak photon-boson diagram (see Figure~\ref{fig:ewbox}) which enters into a number of fundamental processes in subatomic physics.  The $\gamma-Z$ box is a significant contribution
to the asymmetry in parity-violating electron scattering and the $\gamma-W^\pm$ box is an important radiative correction in $\beta-$decay which must be implemented to extract $V_{ud}$ of the Standard Model from $0^+ \rightarrow 0^+$ super-allowed nuclear $\beta$-decays.

The proton form factors are fundamental to hadronic physics.
Understanding the QED expansion, the role of two-photon exchange, and the scale of radiative corrections at higher $Q^2$ will be crucial in future studies at FAIR, JLab, EIC, and elsewhere.  The charge-averaged yield eliminates all charge-odd radiative corrections including the leading terms of two-photon exchange, which cannot be calculated with current theories.  Measuring the ratio of positron-proton to electron-proton scattering is sensitive to the charge-odd radiative corrections and insensitive to the charge-even radiative corrections. Together they help to study radiative corrections and unravel the proton form factors.  TPEX, like OLYMPUS, will provide both these measurements at higher $Q^2$.
    
The discrepancy in the form factor ratio has not been resolved and the role played by two-photon exchange continues to be widely discussed within the nuclear physics community~\cite{EWBox:2017aa, Afanasev:2017gsk, Blau:2017du, NSTAR:2017aa, JPos:2017aa}. Further measurements and theoretical work on the role of two-photon exchange on the proton form factors are clearly needed.  However, measurements at higher $Q^2$ and smaller $\epsilon$, where the discrepancy is clear
and TPE are expected to be larger, are difficult as the cross sections decrease rapidly.  In addition, there are not many laboratories capable of providing both electron and positron beams with sufficient intensity.

The best, and for the foreseeable future only, opportunity is at
DESY. This proposal outlines an experiment that could measure
$R_{2\gamma}$ at $Q^{2}$ up to 4.6~(GeV/c)$^{2}$ or higher, and
$\epsilon$ below 0.1 where the form factor discrepancy is clear
(see Figure~\ref{fig:ratio}). Such an experiment would overlap with the existing OLYMPUS data as a cross-check and would map out the two-photon exchange contribution over a broad range in $Q^{2}$ and $\epsilon$ to provide data to constrain theoretical calculations.

%***+****1****+****2****+****3****+****4****+****5****+****6****+****7****+****8

\section{Proposed Experiment}
\label{sec:exper}

\begin{figure}[htbp]
  \centering
  \includegraphics[width=0.46\textwidth] {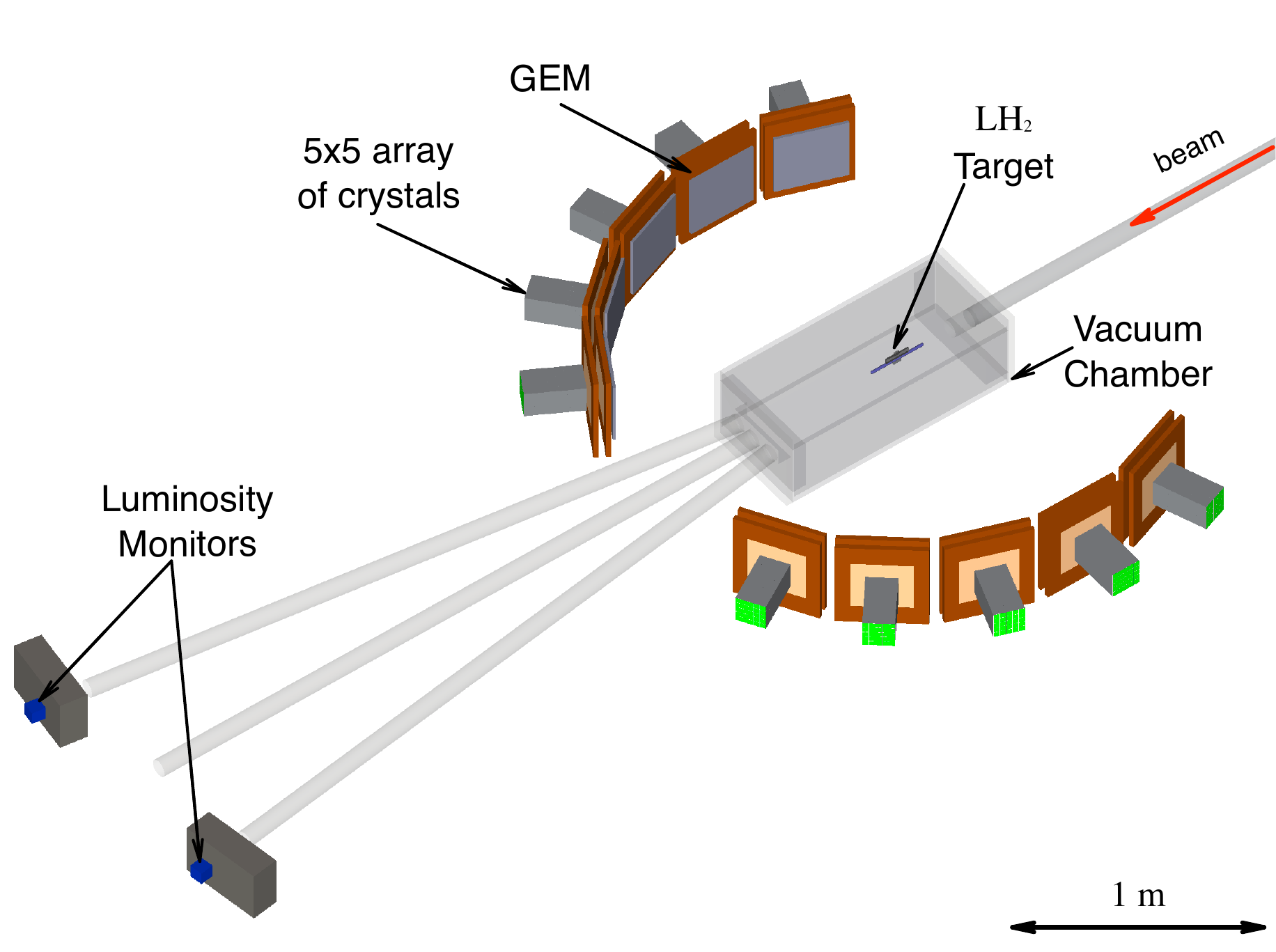}
  \caption{Schematic layout of a proposed TPEX target, scattering chamber, and detector configuration including the luminosity monitors and beamlines.  The lepton beam would enter through the beamline in the upper-right, traverse the target cell, and scatter into the detectors or continue straight to the beamdump.}
  \label{fig:TPEX}
\end{figure}

A schematic overview of the TPEX experimental setup is shown in Figure~\ref{fig:TPEX}.  The electron or positron beam enters the the vacuum chamber along the beamline (upper-right) and passes through the 20~cm long liquid hydrogen target (Section ~\ref{ssec:LH2}) before reaching the beamdump (Section ~\ref{ssec:beamdump}).

At $\pm8\degree$ there are 3~m long beampipes that connect the scattering chamber to the lead collimators before the Cherenkov detectors, used to monitor the luminosity (Section~\ref{ssec:Lumi}). These beamlines are under vacuum and reduce the multiple scattering for the relatively low energy (30--50~MeV) M{\o}ller and Bhabha scattered leptons.  
Ten scattered particle spectrometers are placed at polar angles of $30\degree$, $50\degree$, $70\degree$, $90\degree$, and $110\degree$ to the left and right of the beam axis with the front face of the calorimeter modules at a radius of 1~m from the target. Each spectrometer consists of  a $5\times5$~array of lead tungstate crystals calorimenter (Section~\ref{ssec:cal}) and two planes of GEM detector (Section~\ref{ssec:GEMs}).

%***************************************************************
\subsection{Liquid Hydrogen Target and Scattering Chamber}
\label{ssec:LH2}

\begin{figure}[!ht]
  \centering
  \begin{subfloat}[]{\label{fig:chamber_1}
    \includegraphics[width=0.45\textwidth, clip]
  {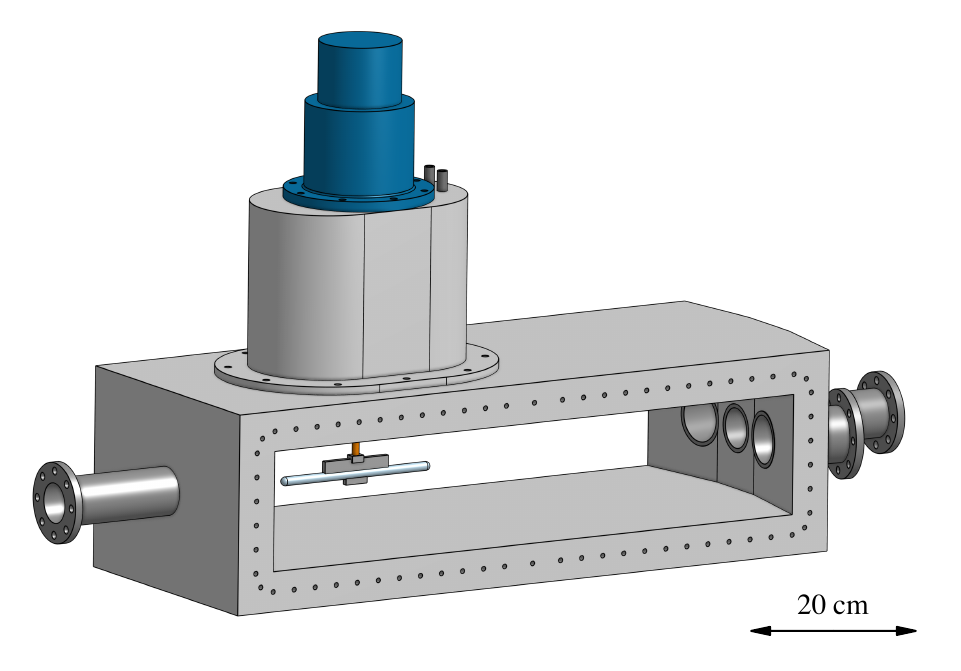}}
  \end{subfloat}
  \hfill
  \begin{subfloat}[]{\label{fig:chamber_2}
    \includegraphics[width=0.25\textwidth,  clip]
  {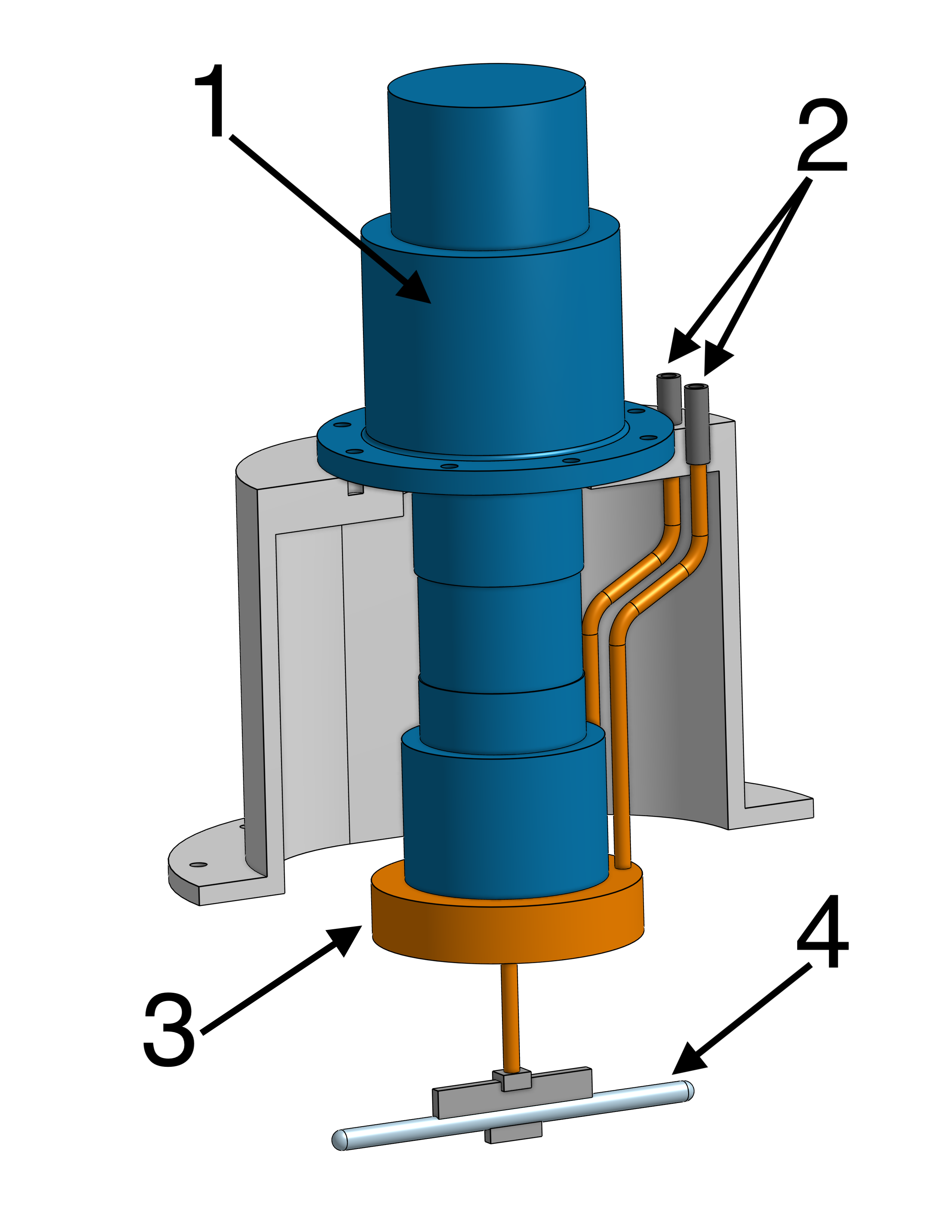}}
  \end{subfloat}
  \caption{Conceptual design of the TPEX target chamber:
  (a) shows the full chamber view with the lepton beam entering from the left; (b) is a sectional drawing of the cryocooler system (1~--~CH110-LT cryocooler, 2~--~hydrogen supply and exhaust lines, 3~--~condenser with a cooling loop, 4~--~target cell).}
  \label{fig:target_chamber_design}
\end{figure}

Figure~\ref{fig:target_chamber_design} depicts the conceptual design of the liquid hydrogen target system.  Figure~\ref{fig:chamber_1} provides a schematic overview of the target system, which consists of the scattering chamber, the cryo-cooler system, and the 20\,cm long, and 2\,cm wide  target cell.  The dimensions of the scattering chamber windows  are determined from the solid angle subtended by the calorimeters. The two side exit windows cover the polar angles for the lead tungstate calorimeters in the range of $\ang{25} < \theta < \ang{120}$. At the end of the 3~m long beam pipes leading to the luminosity monitors are two small exit windows  cover a range of $\ang{7} < \theta < \ang{9}$.  The vertical dimensions of the two side exit windows cover an azimuthal angle of $\phi = \ang{0} \pm \ang{10}$. To maximize rigidity and withstand the enormous force from atmospheric pressure, as well as to avoid welded and bolted joints, we propose to machine the scattering chamber from a single piece of aluminum.

Details of the liquid hydrogen target system are shown in Figure~\ref{fig:chamber_2}.  The cryo-cooler/condenser combination will closely follow the successful MUSE design~\cite{MUSE_Target}. We will therefore use the CH110-LT single-stage cryo-cooler from Sumitomo Heavy Industries Ltd~\cite{CH-100_Manual} for refrigeration. This cryo-cooler has a cooling power of $25$\,W at $20$\,K, which is more than sufficient to cool down and fill the 70 ml LH$_2$ target cell in approximately 2\,hours~\cite{MUSE_Target}.

%%%***********************************************
\subsection{Lead Tungstate Calorimeters}
\label{ssec:cal}

For the proposed experiment, we are capitalizing on the R\&D experience~\cite{Zhu:1996tt,Zhu:2004dg} gained from the CMS experiment and its subsequent applications by the Bonn and Mainz groups at CEBAF~\cite{Neyret:1999tr} and for PANDA~\cite{Albrecht:2015zma}. We plan to utilize ten $5\times5$~arrays of lead tungstate (PbWO$_{4}$) crystals, totaling 250 crystals, some of which may already be available from Mainz. By employing the central $3\times3$~array of crystals within each $5\times5$~array to define the acceptance, we achieve a solid angle of 3.6~msr at each angle. With a 20~cm long liquid hydrogen target, the acceptance range spans $\pm5.7\degree$  in both polar and azimuthal angles. Consequently, the data obtained will be averaged over a small range in $Q^{2}$ and $\epsilon$.

\begin{figure}[!htp]
  \centering
    \includegraphics[width=0.45\textwidth, clip]
  {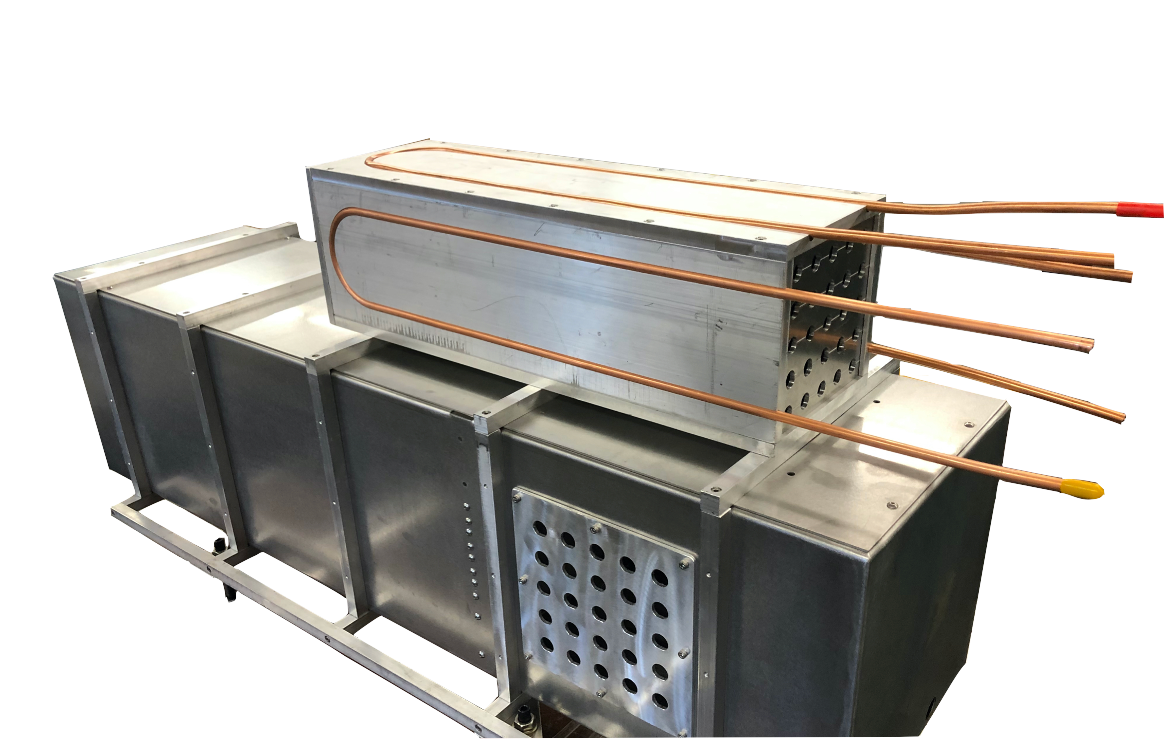}
  \caption{The prototype of the $5\times5$ array of lead tungstate crystals calorimenter tested at DESY.}
  \label{fig:calo_1}
\end{figure}

We plan to use crystals with dimensions of $2~\times~2~\times~20$~cm$^{3}$. With a density of 8.3~g$\cdot$cm$^{-3}$, each crystal weighs approximately 664~g, resulting in a total weight of 16.6~kg for a $5\times5$ array of crystals.  Lead tungstate has a radiation length $X_0=0.8904$~cm, so these crystals are approximately 22.5~$X_{0}$ for good longitudinal electromagnetic shower confinement.  The Moli\`{e}re radius is 1.959~cm, so using just the central $3\times3$~array of crystals for acceptance, the outer ring of crystals contains the transverse shower adequately. Considering the nuclear interaction length of lead tungstate, which is $\lambda_I=20.28$~cm, the crystals are roughly $0.986\ \lambda_I$. For the lepton energy range of interest, the energy resolution achieved with lead tungstate is approximately 2\%. 

The initial tests of the lead tungstate crystals calorimeter prototype (Figure~\ref{fig:calo_1}) at DESY have demonstrated reliable performance and a reconstructed energy resolution that is compatible with the requirements of TPEX. Detailed information regarding these tests can be found in Refs.~\cite{Ivica2023TPEX, TPEX_TDR}.

%***+*******************************

\subsection{GEM Detectors}
\label{ssec:GEMs}

In addition to the lead tungstate crystals calorimenter, each scattered particle spectrometer is equipped with two planes of  Gas Electron Multiplier (GEM) with two-dimensional readout. Thin absorbers will be placed between the target and the GEMs to stop low-energy M{\o}ller or Bhabha leptons. 

The GEMs provide spatial information of the traversing charged particle at the 100 micrometer precision level. By combining the hits on two GEM planes, a track segment is formed, providing directional information from the impact point on the calorimeter back to the event origin in the target. This serves to effectively suppress charged-particle backgrounds originating from regions other than the target. Furthermore, the GEM detectors are insensitive to neutral particles, making them effective in providing a veto against photons and neutrons. Additionally, by incorporating the calorimeter hit as a third tracking point, the efficiency of each component can be measured

An active area of slightly more than 20x20 cm$^2$ is required to fully cover the area of the calorimeter entrance. A total of 20 elements is required to instrument ten calorimeter arms.

\subsection{Luminosity and Beam Alignment Monitor}
\label{ssec:Lumi}

The relative luminosity between the electron and positron running
modes is the crucial normalization for the proposed measurement. The
luminosity could be monitored by a pair of small-angle detectors
positioned downstream on either side of the beamline. This approach
was also used in the OLYMPUS experiment~\cite{Benito:2016cmp}, and
based on the lessons learned from that experiment, could be improved
substantially. Given the running conditions of the proposed
measurement, we favor a pair of quartz Cherenkov counters positioned
$8\degree$ from the beamline to monitor the rates of M{\o}ller and
Bhabha scattering from atomic electrons in the target.

\begin{figure}
    \centering
    \includegraphics[width=0.46\textwidth]{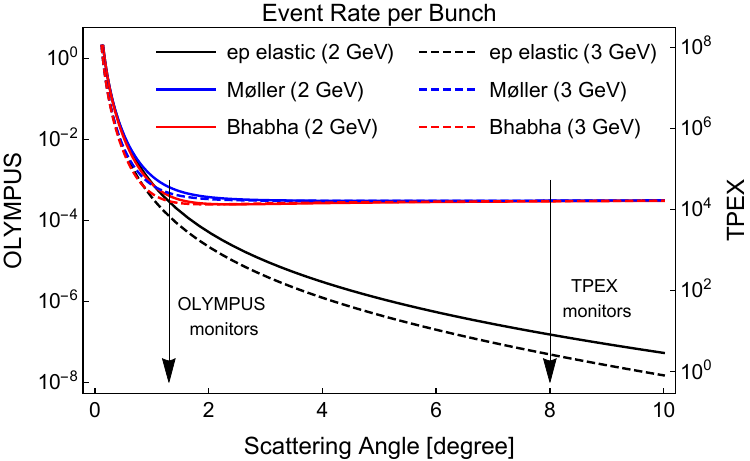}
    \caption{Whereas the forward monitors in OLYMPUS had an event per
      bunch rate well below 1, the TPEX monitors will see $10^4$
      M{\o}ller or Bhabha events per bunch crossing.}
    \label{fig:mb_rates}
\end{figure}

 A monitor placed at $8\degree$ has a number of advantages relative to the $1.3\degree$ placement of the OLYMPUS luminosity monitors.
 First, at $8\degree$, the M{\o}ller and Bhabha cross sections are only a few percent different, whereas for the OLYMPUS monitors, which covered the symmetric angle ($90\degree$ in the center-of-mass frame), the two cross sections differed by over 50\%, with significant angular dependence. Second, the M{\o}ller/Bhabha rate completely dwarfs the $e^\pm p$ elastic scattering rate, meaning that it is really only sensitive to QED processes. No form factors or any other hadronic corrections\footnote{other than the radiative correction from vacuum polarization} are needed to calculate the M{\o}ller and Bhabha cross sections. Third, the sensitivity to alignment scales as $1/\sin\theta$, meaning the monitor will be much more robust to small misalignments, which were a significant challenge for the OLYMPUS luminosity monitor.

\begin{figure}
    \centering
    \includegraphics[width=0.45\textwidth]{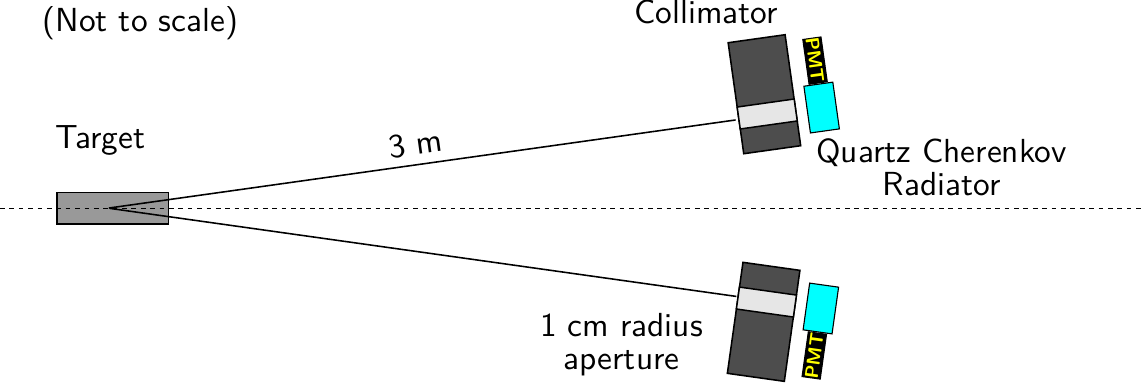}
    \caption{Schematic for the proposed luminosity monitor, consisting of two quartz Cherenkov detectors with an acceptance defined by 1~cm radius apertures in high-Z collimators.}
    \label{fig:lumi_monitor_design}
\end{figure}

 A schematic layout of the design is shown in Figure~\ref{fig:lumi_monitor_design}. The monitor consists of two quartz Cherenkov detectors, which act as independent monitors. Cherenkov detectors were chosen because they are widely used for monitoring in high-rate applications, such as in parity-violating electron scattering~\cite{Abrahamyan:2012gp,Allison:2014tpu}. The two detectors operate independently and can cross-check each other, helping to reduce systematic errors from beam alignment.

%***+****1****+****2****+****3****+****4****+****5****+****6****+****7****+****8

\subsection{Beamdump / Faraday Cup}
\label{ssec:beamdump}

\begin{figure}[h]
    \centering
    \includegraphics[width=\linewidth]{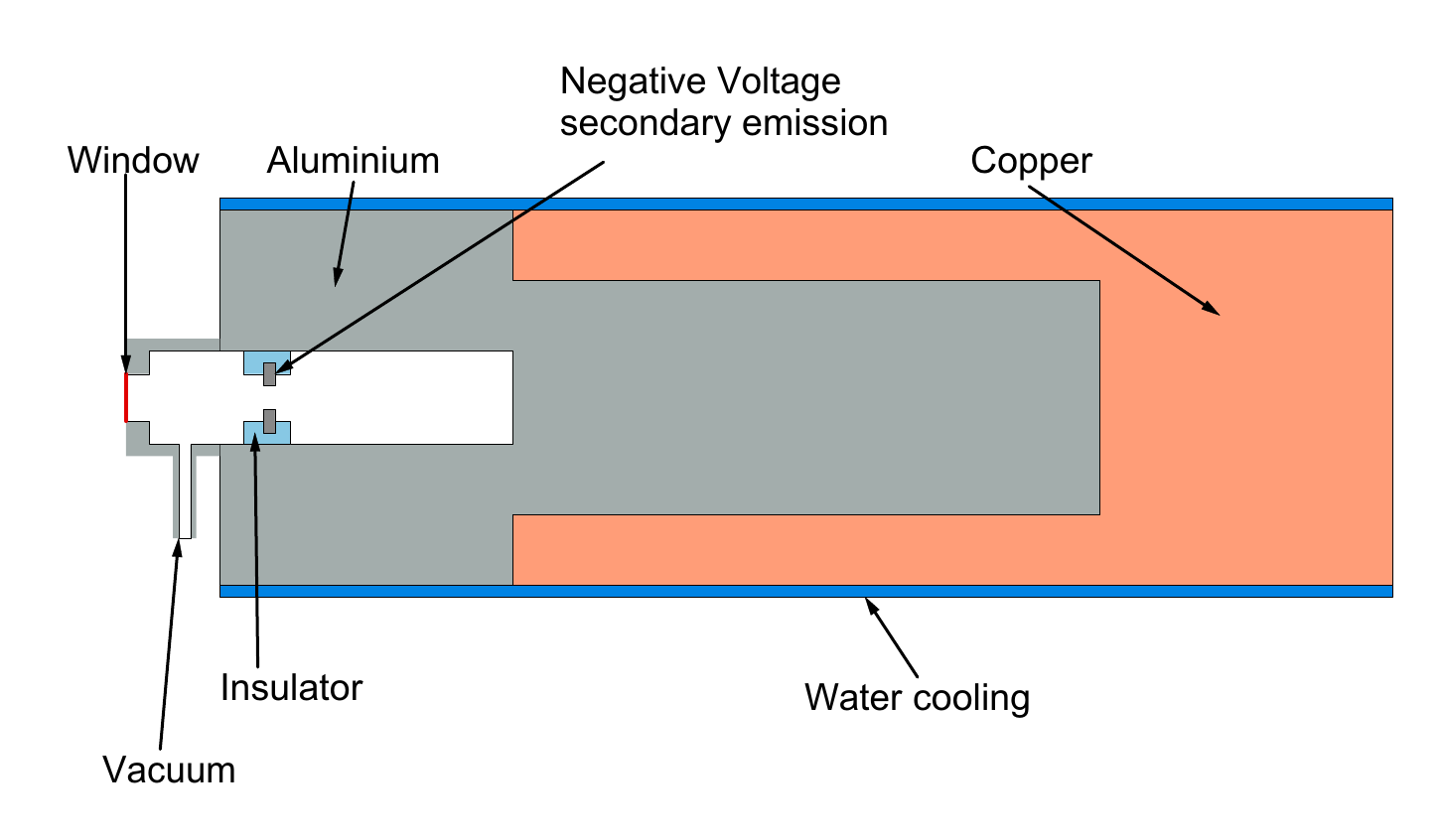}
    \caption{Schematic view of a possible beamdump / Faraday cup for TPEX.}
    \label{fig:beamdump}
\end{figure}

A new extracted beam facility from DESY~II will need a beamdump. Figure~\ref{fig:beamdump} shows the conceptual design of the beamdump for TPEX experiment. Assuming a maximum current of 100~nA and a beam energy of 7~GeV the maximum power to be handled is 700~W.  To contain the showering. the beamdump used to have order of 5~Moli\'{e}re radii laterally and order of 25~radiation lengths longitudinally.

To augment the luminosity measurement proposed in Section~\ref{ssec:Lumi} it is considered to modify the beamdump to function as a Faraday cup as well to integrate the charge that passes through the target.  Then, assuming the length of the target cell and density of liquid hydrogen are known, a quick measure of the luminosity can be obtained.  As shown in Figure~\ref{fig:beamdump}, an insulated ring held at negative voltage of a few hundred volts is needed to suppress secondary emission from back scattering out of the Faraday cup.

%***+****1****+****2****+****3****+****4****+****5****+****6****+****7****+****8

\section{Plans and Expectations}
\label{sec:data}

We propose to commission the experiment using 2~GeV electrons.  We do this to commission the electronics, detectors, and data acquisition system taking advantage of the relatively high cross section at 2~GeV.  About 2 weeks of beam time is required for this commissioning after the experiment was installed and surveyed.  We would also like a brief run (few days) with positrons to verify that the beam alignment and performance do not change with positron running.  The commissioning
run (including a few days with positrons) would also allow a
crosscheck of the OLYMPUS data at $30\degree$, $50\degree$, and
$70\degree$ and give a modest extension in $Q^2$ up to
2.7~(GeV/$c$)$^2$.

Table~\ref{tb:2GeV} shows $Q^{2}$, $\epsilon$, differential cross section, and event rate expected for one day of running for the proposed left/right symmetric configuration with 2~GeV lepton beams averaging 40~nA on a 20~cm liquid hydrogen target and using just the central $3\times3$ array of crystals to calculate the acceptance area.

\begin{table}[htbp]
  \centering
  \begin{tabular*}{0.48\textwidth}{@{\extracolsep{\fill}}ccccc}
    $\theta$&$Q^{2}$&$\epsilon$&$d\sigma/d\Omega$&Events/day\\
            &(GeV/c)$^{2}$&&fb\\
    \hline
    $30\degree$&0.834&0.849&$2.41\times10^{7}$&$3.16\times10^{6}$\\
    $50\degree$&1.62&0.611&$7.66\times10^{5}$&$1.01\times10^{5}$\\
    $70\degree$&2.19&0.386&$1.00\times10^{5}$&$1.32\times10^{4}$\\
    $90\degree$&2.55&0.224&$2.81\times10^{4}$&$3.70\times10^{3}$\\
    $110\degree$&2.78&0.120&$1.22\times10^{4}$&$1.61\times10^{3}$\\
  \end{tabular*}
  \caption{Kinematics, cross section, and events expected in one day
    for an incident lepton beam of 2~GeV and 40~nA averaged current on
    a 20~cm liquid hydrogen target.}
  \label{tb:2GeV}
\end{table}

The TPEX experiment proper would run at 3.0~GeV and would require
approximately 6~weeks (2 weeks with electrons and 4 weeks with
positrons in total) to collect the required statistics.  Table~\ref{tb:3GeV} shows $Q^{2}$, $\epsilon$, differential cross section, and event rate expected for one day of running for the proposed configuration with 3~GeV lepton beams. This would extend the measurements to $Q^2=4.57$~(GeV/$c$)$^2$ where the form factor ratio discrepancy is large. The 6~weeks could be divided into two three-week periods if that was more convenient.  To minimize systematics we would like to switch between positron and electron running as frequently as possible ({\it{e.g.}}  1~day positron, 1~day electron, and 1~day positron repeating).

\begin{table}[htbp]
  \centering
  \begin{tabular*}{0.48\textwidth}{@{\extracolsep{\fill}}ccccc}
    $\theta$&$Q^{2}$&$\epsilon$&$d\sigma/d\Omega$&Events/day\\
            &(GeV/c)$^{2}$&&fb\\
    \hline
    $30\degree$&1.69&0.825&$2.41\times10^{6}$&$3.16\times10^{5}$\\
    $50\degree$&3.00&0.554&$6.51\times10^{4}$&$8.55\times10^{3}$\\
    $70\degree$&3.82&0.329&$8.94\times10^{3}$&$1.17\times10^{3}$\\
    $90\degree$&4.29&0.184&$2.65\times10^{3}$&$3.48\times10^{2}$\\
    $110\degree$&4.57&0.096&$1.20\times10^{3}$&$1.58\times10^{2}$\\
  \end{tabular*}
  \caption{Kinematics, cross section, and events expected in one day
    for an incident lepton beam of 3~GeV and 40~nA averaged current on
    a 20~cm liquid hydrogen target and 3.6~msr acceptance and a
    left/right symmetric detector configuration.}
  \label{tb:3GeV}
\end{table}

\begin{figure}[htbp]
  \centering 
  \includegraphics[width=0.46\textwidth] {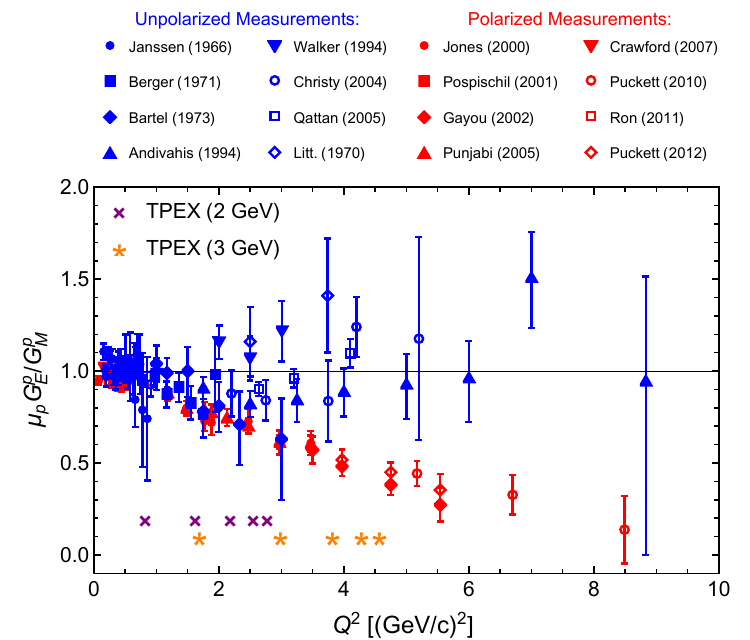}
  \caption{Proton form factor ratio as before but also showing the
    $Q^2$ range accessible with the proposed TPEX configuration at 2
    and 3~GeV. The 4~GeV range would be possible with additional
    crystals.}
  \label{fig:Q2TPEX}
\end{figure}

The $Q^2$ range that the proposed TPEX experiment would be capable of
reaching is shown in Figure~\ref{fig:Q2TPEX} for the 2 and 3~GeV runs of
this proposal. The reach with TPEX can be seen in relation to the
discrepancy in the form factor ratio.  With additional crystals at
back angles the 4~GeV runs would also be possible in a reasonable time
frame.

\begin{figure}[htbp]
  \centering 
  \includegraphics[width=0.46\textwidth] {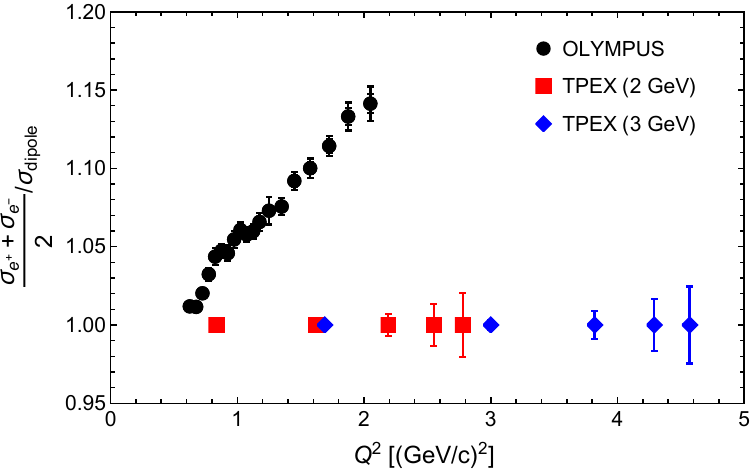}
  \caption{Charge-averaged cross section divided by the dipole cross
    section from OLYMPUS and expected uncertainties and coverage from
    TPEX at 2 and 3~GeV.}
  \label{fig:Qaverage}
\end{figure}

The TPEX experiment at DESY would also measure the charge-averaged
cross section just like the recent result from OLYMPUS (see Figure~\ref{fig:yield}).  As mentioned above this cross section is
insensitive to charge-odd radiative corrections including ``hard''
two-photon exchange terms.  Thus, it provides a more robust measure of
the proton form factors.  The expected charge-averaged cross section
uncertainties (assuming dipole cross section) are shown in Figure~\ref{fig:Qaverage} for TPEX assuming 6 days of running at 2~GeV and 6 weeks of running at 3~GeV with only 50\% data collection efficiency. The recent OLYMPUS results are also shown.

%***+****1****+****2****+****3****+****4****+****5****+****6****+****7****+****8

\section{Conclusion}
\label{sec:concl}

The observed discrepancy in the proton form factor ratio presents a fundamental challenge in nuclear physics and quantum electrodynamics (QED). Despite the inclusion of leading order QED radiative corrections, these corrections alone have proven insufficient to resolve the discrepancy. This suggests that higher order corrections might be necessary to achieve a comprehensive understanding of the phenomenon. Additionally, it is plausible that more detailed models for the intermediate hadronic state could be required to accurately account for the observed deviation. Furthermore, it is crucial to consider the possibility of alternative processes that may contribute to the observed discrepancy.

To address this issue and gain further insights into the proton form factors at higher momentum transfers, the establishment of an extracted positron and electron beam facility at DESY (Deutsches Elektronen-Synchrotron) would offer a unique opportunity. Such a facility would enable the measurement of the two-photon exchange contribution to elastic lepton-proton scattering across a kinematic range where the evident discrepancy is prominent. The proposed TPEX experiment outlines an initial plan for an experimental configuration that could help resolve this issue and provide insight to the radiative corrections needed to understand the proton form factors at higher momentum transfers.

\section{Acknowledgments}
The measurements leading to these results have been performed at the
Test Beam Facility~\cite{DESY_TBL} at DESY Hamburg (Germany), a member of the Helmholtz Association (HGF).
This work was supported by the US National Science Foundation (NSF) grants PHY-2012114, PHY-1812402, PHY-2113436, PHY2012430, PHY2309976, and PHY-2110229, by the US Department of Energy Office of Science, Office of Nuclear Physics, under contract no. DE-SC001658, and grants DE-FG02-94ER40818 and DE-SC0013941. We also received support from the PIER Hamburg-MIT/BOS Seed Project PHM-2019-04 and MEYS of Czech Republic under grant LM2023034.

%***+****1****+****2****+****3****+****4****+****5****+****6****+****7****+****8
\bibliography{sn-bibliography}

%***+****1****+****2****+****3****+****4****+****5****+****6****+****7****+****8

\end{document}